# Ionically gated perovskite solar cell with tunable carbon nanotube interface at thick fullerene electron transporting layer: comparison to gated OPV

*D.S. Saranin[1,2], D.S. Muratov[3], R. Haroldson[4], A. G. Nasibulin[5], A.R. Ishteev[1,3], D.V. Kuznetsov[1,3], M.N. Orlova[2], S.I. Didenko[2] and A.A. Zakhidov[1,4]*

[1]Energy Efficiency Center, National University of Science and Technology MISiS, Moscow, 119049, Russia

[2]Department of Semiconductor Electronics and Device Physics, National University of Science and Technology MISiS, Moscow, 119049, Russia

[3] Department of Functional Nanosystems and High Temperature Materials, National University of Science and Technology MISiS, Moscow, 119049, Russia

[4]Physics Department and The NanoTech Institute, The University of Texas at Dallas, Richardson, 75080, USA

[5]Center for Photonics and Quantum Materials, Skolkovo University of Science and Technology, Moscow,143026, Russia

**Abstract**

We demonstrate an ionically gated planar PS-PV solar cell with ultra-thick fullerene ETL with porous CNT electron collector on top of it. Perovskite photovoltaic devices usually have undoped electron transport layers, usually thin like C60 due to its high resistance. Metallic low work function cathodes are extremely unstable in PS-PV due to reaction with halogens I-/Br-, and it would be desirable to have stable carbon cathodes on top of thick low resistance ETL for enhancing stability of PS-PVs. We show that gating such top CNT cathode in ionic liquid, as part of a supercapacitor charged by Vg tunes the Fermi level of CNT by EDL charging, and causes lowering of barrier at of C60/C70 ETL. Moreover, at higher gating voltage ions further propagates into fullerene by electrochemical n-doping, that increases dramatically PV performance by raising mostly two parameters: Isc and FF, resulting in PCE efficiency raised from 3 % to 11 %. N-doping of ETL strongly enhances charge collection by ETL and CNT raising Isc and lowering series resistance and thus increasing strongly PCE. Surprisingly Voc is not sensitive in PS-PV to external Vg gating, on the contrary to strongly enhanced Voc in ionically gated organic PV, where it is the main gating effect. This insensitivity of Voc to lowering of work function of Vg gated CNT electrode, is a clear indication that Voc in PS-PV is determined by inner p-i-n junction formation in PS itself, via accumulation of its intrinsic mobile ionic species halogens and cations and their vacancies.

**Introduction**

Development of perovskite photovoltaics has started from dye-sensitized solar cells (DSSC) conception in pioneer works[1] by using a metal organic semiconductor as a promising absorber instead of usual dye sensitization. Later, as perovskite solar cells (PSC) began presenting a separate scientific direction of photovoltaics,[2] DSSC structures gradually gave way to mesoscopic and planar configurations[3,4] due to advantages of perovskite semiconductor properties in thin film device application: ambipolar transport,[5–7] suppressed recombination,[8–10] big diffusion length,[11] direct[12,13] and easily tunable band gap.[14,15]

Record PCE of more than 25.2%[16] was demonstrated for perovskite solar cells fabricated on *n-i-p* mesoscopic architectures, which require mesoporous and compact $TiO_2$ film. Despite the fact that such a device structural concept was realized in large-scale printing techniques of DSSC and perovskite *n-i-p* modules,[17–19] this configuration limits the flexible lightweight device application due to the high temperature sintering process needed for TiO$_2$ (up to 450°C). Hole transport layers for most high-performing n-i-p devices were fabricated with small molecule Spiro-Ometad (2,2',7,7'-Tetrakis[N,N-di(4-methoxyphenyl)amino]-9,9'-spirobifluorene),[20] which should be p-doped to increase the life time[21] and enhance the conductivity of the hole.[22] PSC can display high PCE with various Spiro-Ometad HTL thicknesses ranging from the common 200 nm to an anomalistic 600 nm, depending on the doping level to avoid surface recombination, form high open-circuit voltage and use rough perovskite absorber films for higher level of absorption[23]. On the other hand, doping impurities (widely used Li-TFSI, tBP) are the main reason for interface degradation and device operation –instability,[24] such as accelerated photo oxidation.

Another interface problem is from the perovskite side, which has high defect density at interfaces due to accumulation of mobile iodine vacancies, excessive ions,[25] and ferroelectric poling with slow polarization.[26] One promising effort for perovskite interface stabilization is the use of fullerene based passivation layers[27]—C60, C70, PCBM, and others. It was shown[27] that fullerene passivation can effectively increase the surface recombination lifetime with total decreasing of surface trap density by capping perovskite grain boundaries and C60 penetrating into the bulk between grains. For p-i-n (inverted) PCS, the fullerene thickness range for device operation feasibility lays is between extremely thin 2.5 nm to the commonly used 50 nm[28]. Devices with C60 ETL thicknesses below 25 nm usually show low shunt resistance with a reduced short circuit current and filling factor. Fullerene layer with thicknesses of 30–50 nm lift the overall PCE to high performance level, but the concept of a thick, robust, passivation electron transport layer with suitable conductivity as analogue to doped Spiro-Ometad HTLs is highly desirable but not yet presented, However, it can be potentially realized with co-evaporation techniques using n- type dopants like PhIm, acrydine orange, CoCp2, etc,[29–31] similar to OLED technology but these methods require expensive materials and complicated technological processes.

Electrode interface also suffers from instability factors: metal diffusion into transport layers and perovskite absorber,[32] appearance of bubbles in metal films during evaporation,[33] and halogenic oxidation.[34] Most prospects for metal replacement focus around carbon because this material can be printed or laminated without vacuum processes, stabilize the interface, and demonstrates hydrophobic properties. Currently, most stable PSC are fabricated with carbon electrodes;[35] moreover, carbon can be effectively used for semitransparent solar cells in forming nanotubes. Several papers show competitive results of cells with single-walled CNT anodes,[36,37] but this material is still being pursued for metal conductivity and $W_f$ engineering.

A promising effort for improving the electrode-transport layer junction is interface engineering and accumulation of charge carriers. Electrode and transport layer energy levels can be aligned via interfacial polarization and dipoles caused by a buffer layer based on ionic liquids. As shown in works with organic solar cells by Yu,[38] Zhang,[39] and Kang,[40] using ionic liquids at the cathode interface improved electron collection with metal oxide and polymer transport layers by reducing electrode $W_f$ and forming an ohmic contact. Ionic liquids (IL) are room temperature melting salts usually presented by long chain organic cations and inorganic anions such as BF$_4$, Cl, PF$_6$, and TCA with a wide electrochemical window of stability. Using ILs does not require vacuum deposition of insulators and can be processed with solution techniques. The simplification of fabrication processes has a wide range of approaches in energy related, thin film devices with liquid-solid interfaces.

The gate field effect is another way to improve charge collection or accumulation in thin film solar cells. Cook[41] and co-workers have shown tunable OPV with ionically gated CNT cathode electrode, in which Fermi level of CNT raised significantly, decreasing w.f. by 0.5 eV by negative gate voltage Vg ~ 1.5 V. Causing dramatic increase of all parameters of OPV: Voc from 0.1 V to 0.6 V, FF raising to 0.7 and resulting in significant PCE increase. Zhou and co-authors[42] designed and fabricated OPV cells with a cathode interface gate with significant improvement in charge extraction (+24 % to $J_{sc}$) under gate bias. A more advanced concept was presented in a paper by Karak[41], where electron extraction was achieved in the AC field with an ionic liquid cathode interfacial layer, Additionally, charge accumulation is an effective solution for electro physical property tuning of transport semiconductors. This area of study is the main problem in developing thin film FET (field effect transistors) for modern electronics based on organic semiconductors, metal oxides and carbon derivatives (fullerenes, nanotubes, graphene). Ionic liquids can form a double electric layer (EDL) with applied bias and act as a gate due their specific capacity (can reach more than $\mu F/cm^2$ versus $nF/cm^2$ order in $SiO_2$), which influences the quantity of induced carriers at low gate voltages <5 V [43]. Ionic gate is presented in various types of super capacitors, fuel cells, dye sensitized solar cells, and field effect related devices. And recently ionic gating was applied even in OLED and OLET devices, making them brighter and more efficient by enhanced charge injection at gated interface[44].

In this work, we found motivation through a combined approach of interface engineering with ionic liquid and n- type accumulation of charge carriers by ionic gating under Vg applied to gate. Carbon nanotubes as a high-specific surface network is an advantageous concept for ionic gate. High surface of materials gives possibility to induce higher concentration of carriers in electrostatic regime. In this case, carbon network can act as effective charge plate for a supercapacitor based on ionic liquid. In its turn, such gate concept can be effectively used in three-interface interaction structure, when supercapacitor is placed on semiconductor surface and ionic liquid penetrates through CNT porous web to semiconductor surface and then to its bulk by diffusion/drift.

Herein, we present a novel type of inverted perovskite planar cell with ultra-thickness (up to 300 nm), a robust fullerene electron transport layer (ETL), and a tunable cathode interface based on highly porous carbon nanotubes sheets. We have developed a horizontal architecture, then fabricated and characterized the p-i-n structure with horizontal CNT ionic gate placed at back electrode. In this device, high specific resistance of the undoped ETL layer and the series resistance on the electrode interface are reduced by charge carrier accumulation regimes. The use of the ionic gate is demonstrated as an effective, low voltage tool for improving solar cell output parameters in reversible electrostatic regimes. We compared device structures with two fullerene ETLs–C60 and C70 ranging from 200–300 nm thicknesses at different gate voltages finding correlations to output PV parameters.

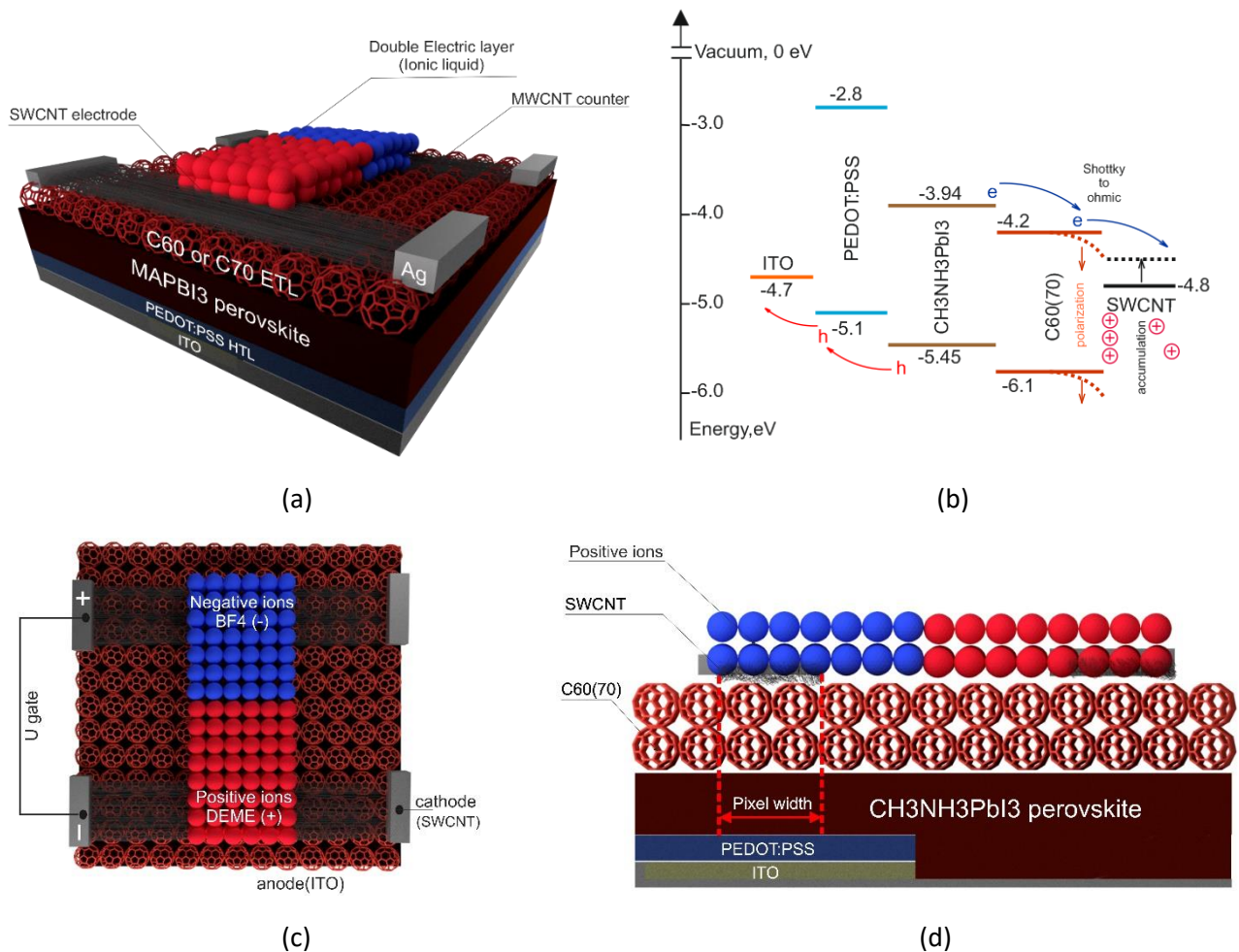

Figure 1. (a) Device schematics of inverted planar MAPbI$_3$ cell with ionic gate between SWCNT cathode and MWCNT electrode, (b) Device band diagram,(c) Device top view,(d) device cross side view

**Experimental part**

**Ink preparation**

Perovskite ink was fabricated in 1.5 M concentration from methylamine iodide (MAI, Dyesol) and lead iodide (Alfa Aesar). Firstly, MAI was dissolved in mixture of DMF: γ-BL (1:1 volume ratio, from Sigma Aldrich, anhydrous), and then PbI$_2$ was dissolved in MAI solution with a 10% volume addition of DMSO. Ink was heated overnight at 60 °C and cooled to room temperature in 5 minutes before device fabrication. PEDOT: PSS water dispersion from Hereaus Clevious (1,4 % Al 4083) was filtrated through 0.45 um PTFE filter prior HTL spin coating.

**Device fabrication**

Pixelated ITO substrates (Lumtec, 15 ohm/□) were cleaned in an ultrasonic bath with acetone, toluene, and IPA (ME grade). Substrates were preliminary activated by 20 minutes UV-ozone treatment for PEDOT: PSS HTL deposition. A 30 nm layer was spin-coated at 3000 RPM during 60s and then annealed at 150 °C. A CH$_3$NH$_3$PbI$_3$ 450 nm photoactive layer was formed after a 2-step spin coating process: 20 s at 1000 RPM and 25 s at 4000 RPM and toluene dripping procedure (10s before 2nd step ending) with final annealing at 100°C during 10 minutes (all processes were provided in glovebox with inert atmosphere < 1 ppm O$_2$, <1 ppm H$_2$0). C60 and C70 electron transport layers (200,250,300 nm thicknesses) were thermally evaporated at 2*10-6 Torr with 0.5 A/s rate.

SWCNTs (synthesized accordingly, route described in previous work) were laminated as 3 mm stripe on the top of device structure and densified with HFE.

MWCNTs (synthesized accordingly, route described in previous work) were laminated as 3 mm stripe on the top of device structure and densified with HFE.

**Ionic gate deposition**

MWCNT as counter electrode was laminated (5 layers) in parallel to the SWCNT cathode in 3 mm distance and densified with HFE.

Finally, a drop (2 μl) of DEME -BF$_4$ ionic liquid (Kanto Chemicals) was squeezed between the SWCNT and MWCNT electrodes by glass coverslip on top of the warm wet planar layer.

Preliminary, 50 and 100 nm ETL devices were tested, but ionic liquid interpenetrated the perovskite layer under pressure of cover slip and partially dissolved photoactive film. For the 50 nm, the ETL layer device degraded in first minute of testing; for 100 nm, the ETL device started to degrade after 5 minutes of measurement and showed poor performance (presented in supplement material). Therefore, thick fullerene film also acts as a protection layer for perovskite in devices with ionic gate.

**Characterization**

JV characterization were provided at standard 1.5 AM G 100 mW/cm2 conditions with Newport ABB solar simulator (calibrated with Si certified cell) and two Keithley 2400 SMU (JV sweep, gate applying) in inert atmosphere.

**Ionic gate JV Sweeps**

Quantum efficiency measurements were provided on XP-6 system (PV MEASUREMENTS) calibrated with KG-9 Si reference cell.

Raman characterization was done on Thermo DXR Raman microscope with 532 nm laser.

## Results

The output performance of fabricated cells was measured in two regimes: without ionic liquid in ITO/HTL/Perovskite/ETL/SWCNT device structure and with SWCNT/DEME-BF4/MWCNT ionic gate at different biases. Gate voltage was applied with positive connection to the counter electrode and negative to the SWCNT device cathode for respective inducing of $DEME^+$ ions at SWCNT-ETL interface and $BF_4^-$ ions at MWCNT counter.

The presence of cation and anion induce polar charge accumulation (electrons and holes respectively) at semiconductor interface provided a doping effect and Fermi level shift in electrostatic regime. As described in our previous works[45,46] in OPV devices, high concentration of cations at carbon nanotube-acceptor interfaces induced by gate voltage raises the Fermi level, as shown in the band diagram of Figure 2(a). At initial conditions, JV curves of cells without ionic liquid gate show a strong S-shaped character caused by high contact resistance between SWCNT cathode and thick, intrinsic C60 and C70 ETLs, as shown in Figures 2 and 5(g–h). For both fullerene materials, such ETL thicknesses between 200–300 nm affect device output performance with series resistance values from ~2000 Ohm*cm$^2$ (C60, C70 200 nm thick) to >12000 Ohm*cm$^2$ (C60, 300 nm) and >7500 Ohm*cm$^2$ (C70, 300 nm) due to the intrinsic low conductivity >$10^{13}$ $\Omega$*cm[47]. Accordingly, typical thicknesses of fullerene based ETLs in perovskite solar cells range 15–50 nm. Moreover, sheet resistance of SWCNTs[48] used for electrodes did not reach the level of typical metal electrodes and conductive oxides[49]. As a result, the devices' FF values are below 0.25, less than that for a resistor-like curve. JV behavior changes dramatically with appearance of SWCNT cathode-ionic liquid interface and n-type accumulation accordingly to gate bias value. Firstly, initial contact between DEME-BF4 with SWCNT cathode without applied bias has dropped series resistance for both fullerene ETL to more than order for 200,250 nm thicknesses and >5 times for 300 nm thick C60 and C70 ETLs cells. Sequential increasing of gate bias from 0.00 V to 1.00 V reduced series resistance to acceptable values of <80 ohm*cm$^2$ for C60 and <40 ohm*cm$^2$ for C70 ETL devices and transferred s-shaped curve to diode-like JV with 0.40–0.50 FF values. Next slight $V_{gate}$ increasing with 0.25 V step (from 1.00 V to 2.50 $V_{gate}$) showed same trend in $R_s$ reducing to level below 20 Ohms*cm$^2$. The highest FF values were obtained at gate biases >1.50 V in range 0.6-0.7 for 200–300 nm C60 samples and 0.5–0.6 for 200–300 nm C70 samples at $V_{gate}$ >1.75 V. A big difference between C60 and C70 ETL devices was obtained in current generation behavior with dependence to thickness of layers. Output performance of cells differs in dynamics of photocurrent increments with increase of $V_{gate}$ bias. This distinction is clearly observed in Figures 5(a, b), where $J_{sc}$ values are shown like function of $V_{gate}$. Spread of $J_{sc}$ gain for C60 ETL devices have more expressed dependence to ETL thickness with increasing of $V_{gate}$ than for C70 cells. We have found that $J_{sc}$ gate response is different at range of bias $\leq$1.00 V; 1.00 > $\leq$2.00 V; and >2.00 V for different thicknesses. The 200 nm ETL devices have not shown significant gain of $J_{sc}$ with increasing gate biases from 0.00 V t0 1.00, for both fullerene types. The C60 ETL cell had $J_{sc}$ growth only +0.56 mA/cm$^2$ from 13.11 mA/cm$^2$, while the C70 ETL cell had gain of +0.60 mA/cm$^2$ from 18.35 mA/cm$^2$ (<4 % in both cases). On the other hand, relative contribution of ionic gating for $J_{sc}$ growth with thicker ETL is much higher, +34.8% and +41.3% for 250 nm and 300 nm C60 films; +31.9 % and +25.3 % for C70 films respectively. At higher gate bias range 1.00 >

≤2.00 V, C70 ETL devices have approximately the same dynamics of $J_{sc}$ increment, as response to the increase of $V_{gate}$ with all used thicknesses, and much lower spread of values in comparison to C60 ETL cells in dependence to thicknesses. At gate voltages >2.00 V, no meaningful $J_{sc}$ was observed. Total improvement in output performance has impressive indicators. The use of ionic gate, as a tool for tuning cathode-ETL interface, increased the PCE of devices more than five times, on average. As shown in Figures 5(e, f), C60 samples improvement is 2.46% to 11.28 % (200 nm); 2.05% to 8.67 % (250 nm); 0.72% to 6.55 % (300 nm), and C70 is 3.35% to 9.74 % (200 nm); 1.47% to 10.74 % (250 nm); 0.72% to 10.31 % (300 nm).

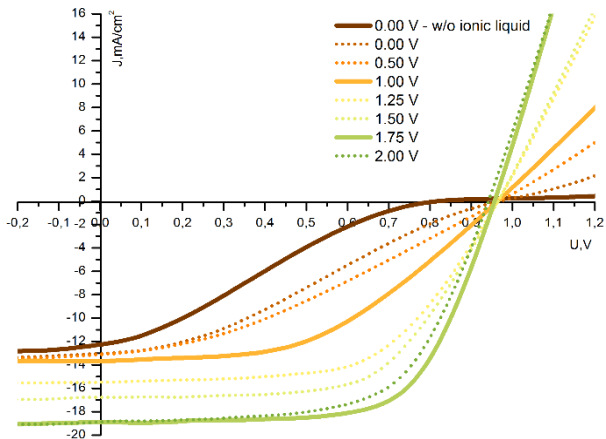

(a) Output performance improvement of 200 nm thick C60 cell from S – shaped JV curve to diode-like characteristics at 0…1.00 V gate with sequential power increase up to 1.75 Vgate

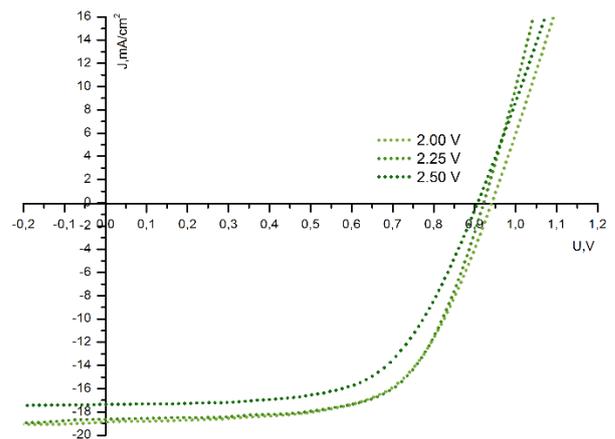

(b) Output performance of 200 nm thick C60 cell at Vgate>2.00 V with slightly decreasing of $J_{sc}$

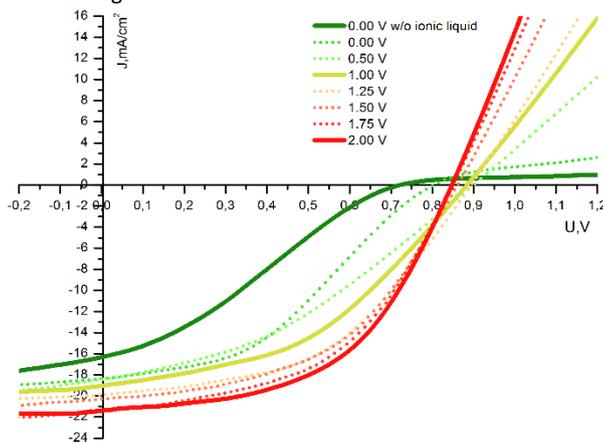

(c) Output performance improvement of 200 nm thick C70 cell from S – shaped JV curve to diode-like characteristics at 0…1.00 V gate with sequential power increase up to 2.00 Vgate

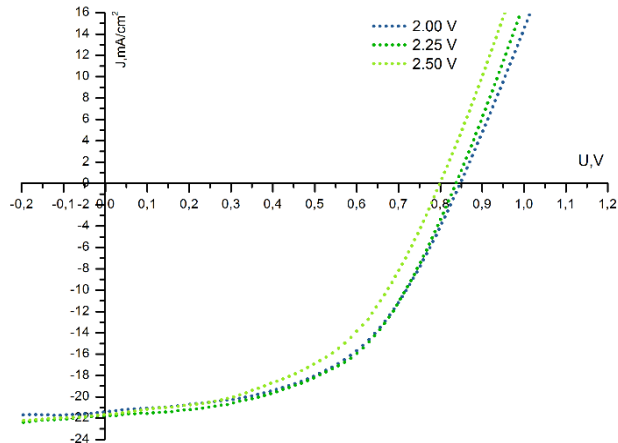

(d) Output performance 200 nm thick of C70 cell at Vgate>2.00 V with slightly decreasing of $V_{oc}$

Figure 2. JV curves of MAPBI$_3$ cell with 200 nm thick C60 (a)(b) and C70 (c)(d) ETL 0 to…2.50 V range of gate bias

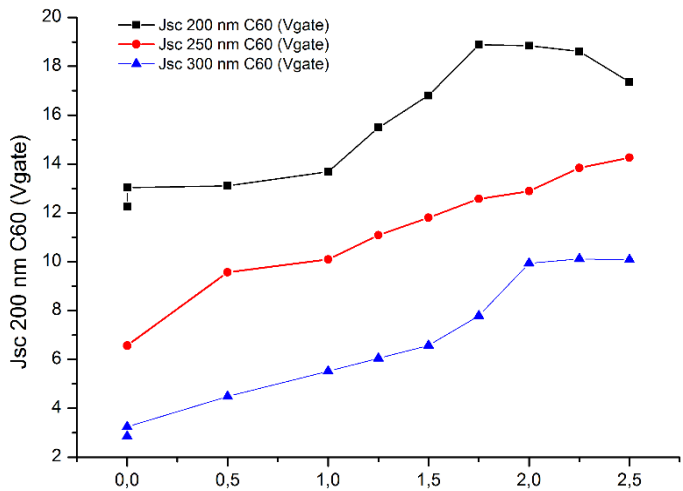

(a) Short current density V gate dependence of 200-300 nm thick C60 cell

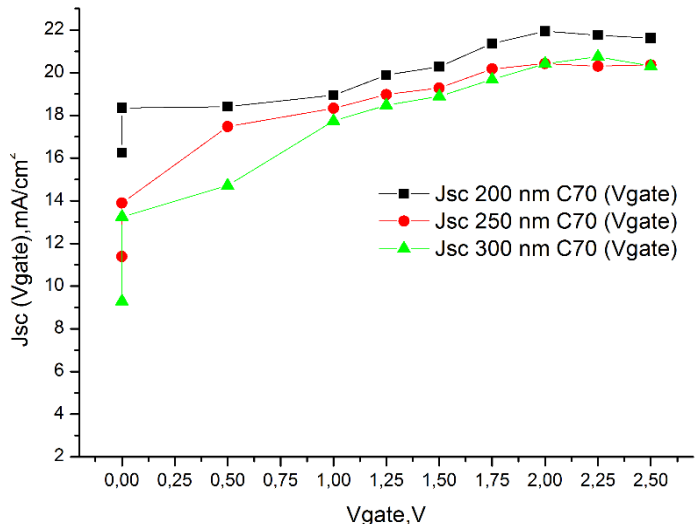

(b) Short current density V gate dependence of 200-300 nm thick C70 cell

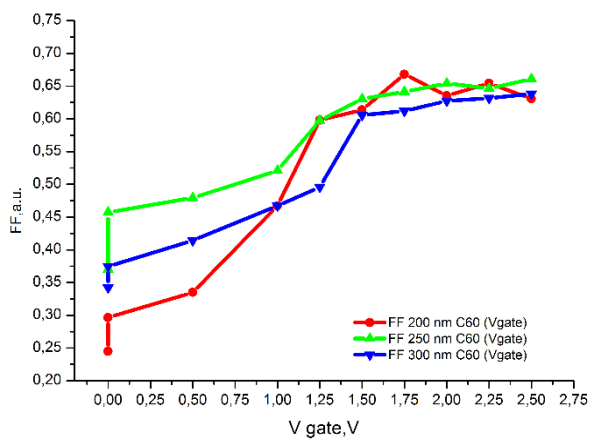

(c) Filling factor V gate dependence of 200-300 nm thick C60 cell

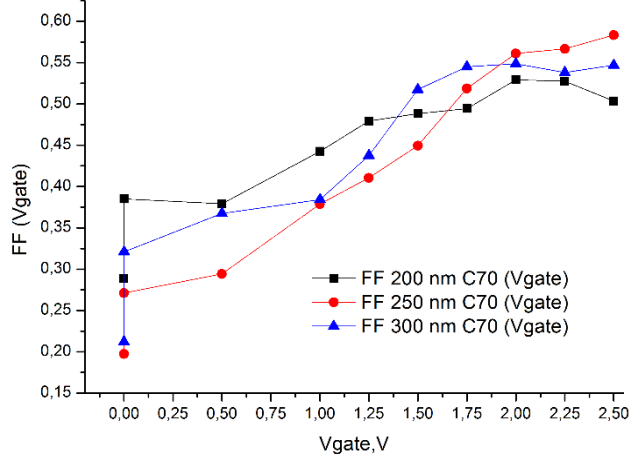

(d) Filling factor V gate dependence of 200-300 nm thick C70 cell

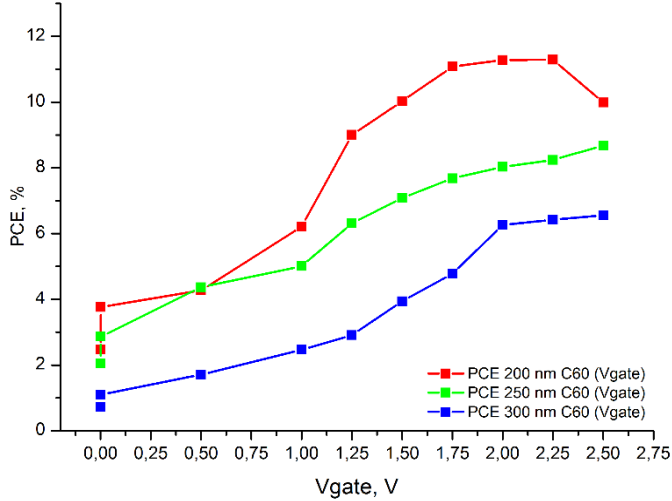

(e) Power conversation efficiency V gate dependence of 200-300 nm thick C60 cell

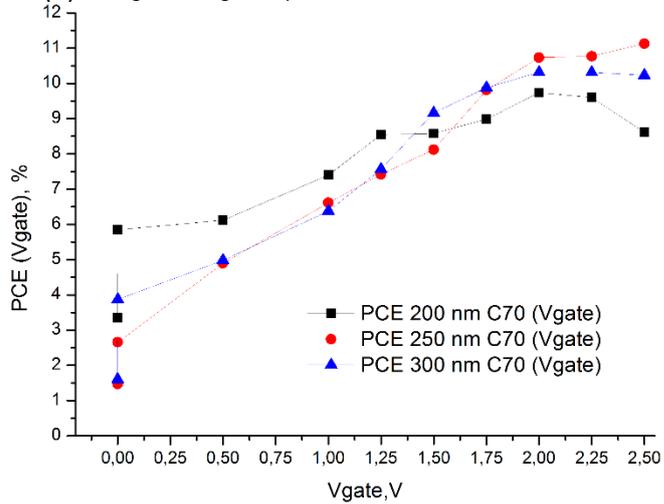

(f) Short current density V gate dependence of 200-300 nm thick C70 cell

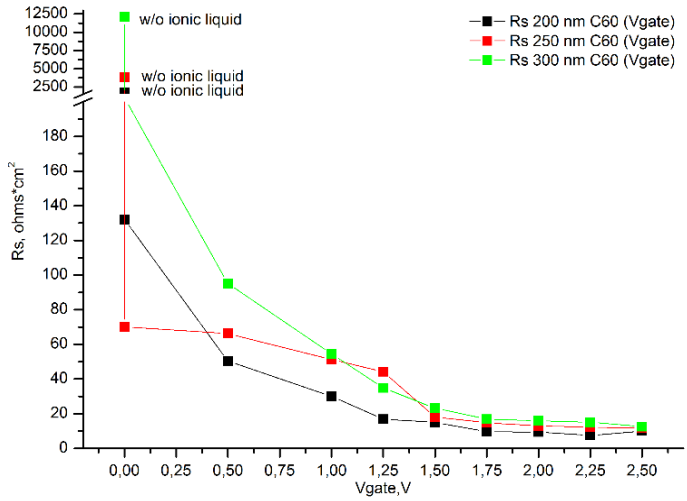
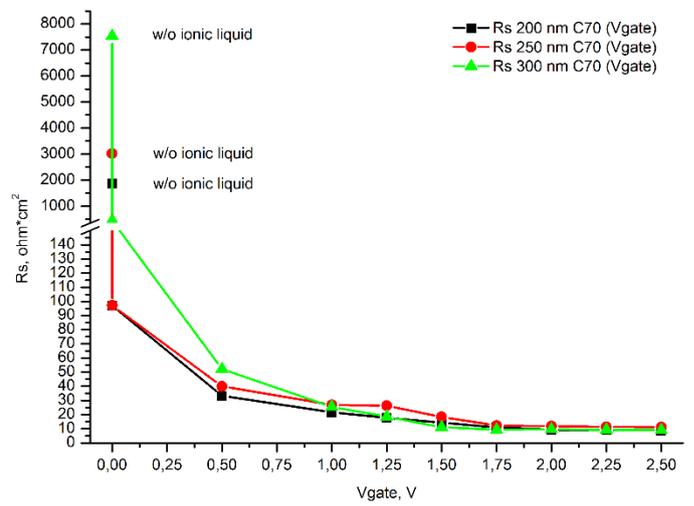

(g) Series resistance V gate dependence of 200-300 nm thick C60 cell

(h) Series resistance V gate dependence of 200-300 nm thick C70 cell

Figure 3. JV Output parameters V gate dependence versus ETL thickness

During quantum efficiency measurements in ionic gating regimes, we obtained the same trend of device performance improvement. Quantum efficiency spectrums have stable growth with increasing $V_{gate}$ for both types of fullerene films and thicknesses (external quantum efficiency spectra of 200 nm C60 and C70 ETL devices presented in Fig. 4, spectra for 250 and 300 nm presented in supplementary). EQE spectra are typical for MAPbI$_3$ perovskite solar cells with PEDOT: PSS and C60-C70 transport layers. Considerable changes are clearly observed in increment of photon quantity converted to electricity. The level of photon conversion grows with same dynamics as short current density during gating in JV testing. As is presented in Figure 4(a), EQE spectra shift upward with bigger gain at 1.00–2.00 V gate bias for the C60 sample and have less response to gate voltage for the C70 cell (Figure 4(b)). In the maximum point of spectra for 200 nm C70 cell, EQE was achieved at 80%, and that is 40% higher than cells at initial conditions without ionic liquid gate, respectively for C60 ETL devices, maximum was achieved at ~65%. The level of photon conversion decreases more than 5–10% at $V_{gate}$>2.00 V [Figure 4(b), 5(b)] with the appearance of QE shoulder in near UV region; it corresponds to perovskite absorber degradation and will be discussed in the next chapter. In general, growth of the EQE spectra follows the trend of $J_{sc}$ changes of solar cells during gating, and confirms improvement of charge collection.

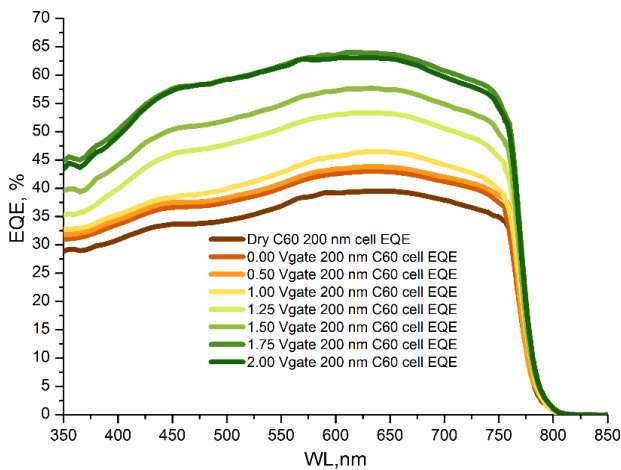
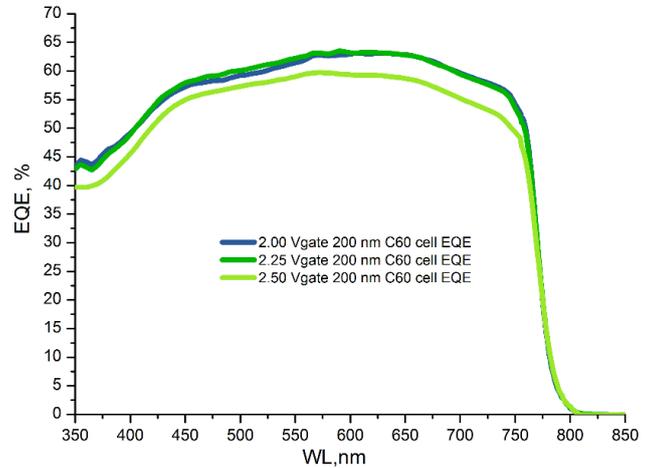

(a) Growth of light conversation EQE level from 350 to 850 nm wavelength at ≤ 2.00 V (200 nm thick C60 cell)

(b) Changes of light conversation EQE level from 350 to 850 nm wavelength at ≥ 2.00 V (200 nm thick C60 cell)

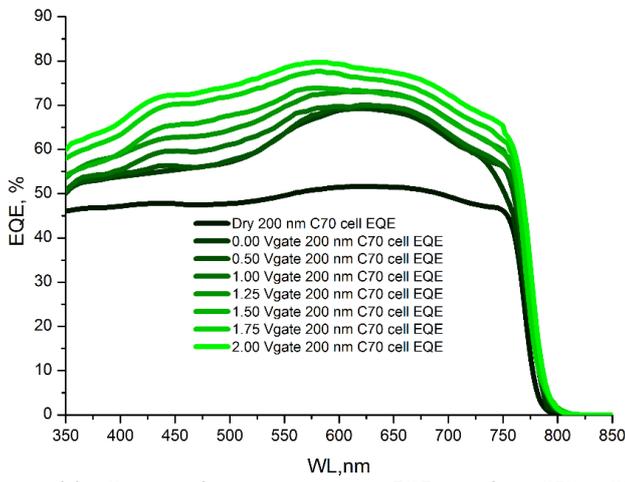
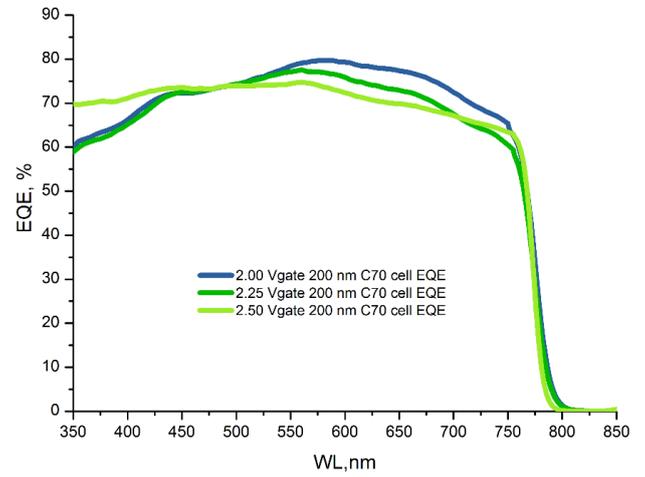

(c) Growth of light conversation EQE level from 350 to 850 nm wavelength at ≤ 2.00 V (200 nm thick C70 cell)

(d) Changes of light conversation EQE level from 350 to 850 nm wavelength at ≥ 2.00 V (200 nm thick C70 cell)

Figure 4. External quantum efficiency spectrums of MAPBI$_3$ cell with 200 nm thick C60 (a)(b) and C70 (c)(d) ETL at 0…2.50 V gate bias

Analysis of fullerene-SWCNT-ionic liquid interface operation was done in separate structures for JV measurement of semiconductor junction. ITO/C60 (C70)/SWCNT-DEME-BF4-MWCNT samples were fabricated in the same route as layers in solar cells, respectively. As presented in Figures 6(a, b) for 200 nm C60 and C70 films, initial resistance of SWCNT-fullerene junction is two orders more than resistance at 2.5 V of gate voltage at ionic liquid gate. This means that n-type accumulation regime significantly increased the ohmic of electrode-semiconductor contacts and provided energy levels matching between SWCNT work function and fullerene LUMO. In addition, JV curve behavior changed two times during the ionic gating process. Initial conditions without an ionic liquid JV curve had a resistor-like type of an approximate straight line (accordingly to Ohm law). Then, asymmetry of forward (V>0) and reverse JV curves (V<0) appeared during increasing gate voltage. The slope of the forward curve grew much higher in comparison to the reverse curve, which, in opposition, moved to axis line. For both ETL materials, the maximum relative between high current values of forward curve and minimum values of reverse curve were achieved at 1.5 gate bias. Valve ratios were calculated as ~10 for 200 nm C60 film and ~4 for C70 respectively. Moreover, until 1.5 V at gate, reverse current decreased dynamics or tendency to leakage removing. We suggest that such JV behavior changes result from the i-n junction appearing in accumulation regime under electrostatic gating. At gate biases >1.50 V, reverse curves swiftly gained the slope and formed a resistor-like JV graph with much smaller differential resistivity in comparison to initial conditions. We believe this effect is consequently due to inducing the n-type doping for the whole depth of ETL. Decreasing of Voc in devices with occurs due solvation of fullerene ETLs in IL in analogue to effect described in work of Maciel[50] and have long-term time dependence.

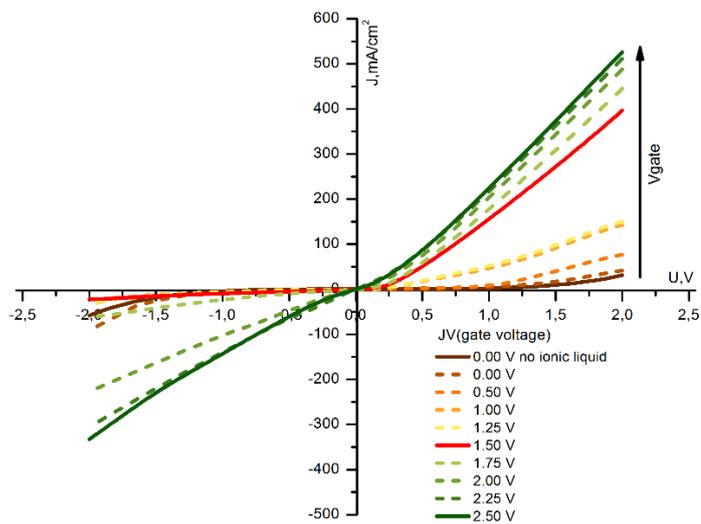 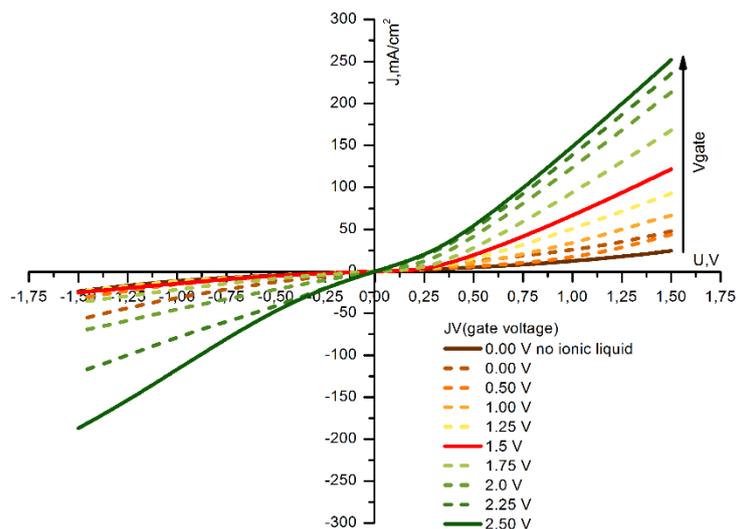

(a) JV behavior changes from resistor (till 1.0 V gate) to diode like curve (at 1.5 Vgate) of ITO/C60 200 nm/SWCNT-IL gate structure with sequential reverse to resistor characteristic with lower differential resistance

(b) JV behavior changes from resistor (till 1.0 V gate) to diode like curve (at 1.5 Vgate) of ITO/C70 200 nm/SWCNT-IL gate structure with sequential reverse to resistor characteristic with lower differential resistance

Figure 5. C60 (a) and C70 (b) diode characteristics with gate bias dependence

**Discussions**

Large progress in PSC performance improvement was achieved with development of doping techniques of thick polymer hole transport layers like Spiro-Ometad, P3HT, PTAA, etc., but not many works tended to analogue approaches with organic electron transport materials. In this work, we show that thick fullerene layers >200 nm can be used as electron transport layers for perovskite solar cells without complicated doping methods like co-evaporation. The use of ionic gate in advanced planar device architecture compensates initial low conductivity of thick ETL and non-metal electrodes in the accumulation charging regime. Electrostatic inducing of n-type carriers at ETL-cathode interface improved mismatching energy levels and provided a significant gain of charge collection in initially intrinsic semiconductors (C60/C70, SWCNT).

The use of horizontal ionic gate was addressed for managing three critical points in devices with thick ETL and undoped SWCNT electrodes: first, high sheet resistance of cathode; second, high resistance of thick ETL; and third, not optimal fullerene LUMO-SWCNT $W_f$ matching. For optimal electron collection and provision of ohmic contact, zero barrier between the transport layer and electrode is required. The measurement of the SWCNT's work function is imprecise due containment of metallic and semiconductor parts and approximate acceptance equal to 4.7 eV, according to data from the literature[51]. In this case, initially the ETL-cathode junction has 0.4 eV of energy loss caused by mismatching of ETL LUMO and electrode $W_f$. On the other side, high ohmic fullerene films cannot provide high electric current flow with 60 Ohm/☐ SWCNT, and this significantly reduces the slope of the IV curve. Such affection was clearly observed in JV behavior of tested solar cells and diodes in structures without ionic gate. Therefore, formulation of problem for junction was defined as necessity of SWCNT $E_f$ raise and increasing of local conductivity in ETL via higher concentration of charge carriers (n-type doping in both cases). According to this statement, the device's operation mode should provide accumulation of positive ions (cations) to induce carriers with opposite charge at back electrode interface. During negative charge accumulation at SWCNT-DEME$^+_{cation}$, the interface Fermi level of SWCNT starts to shift up, providing better work function and matching with LUMO of fullerene. In turn, the ETL surface is also working under gate accumulation regime via soaking by ionic liquid through CNT network, and near-surface layers of fullerenes become n- doped in electrostatic mode. The level of accumulation was controlled by gate voltage (0.00–2.50 V), and in turn, high concentrations of induced carriers are provided via the high capacity of ionic liquid[43] and the high specific area of the SWCNT acting as mediators for positive ions. The charging process of ion induction and distribution on the SWCNT cathode and MWCNT counter was determined as a time dependent process between condenser plates. Cathode-counter current flow has an initial level of pA without gate voltage and falls with hyperbolic dependence from ~100 nA to saturation at ~1 nA after applying the gate bias. Therefore, such additional power losses (~nW) for ionic charging has no critical contribution to the device operation. Typical charging I(t) plots presented in supplement material.

There is a very significant qualitative difference in the effect of Vg on Voc of perovskite PV as compared to our earlier results on ionic gating in Polymeric PV[52] and small molecule PV [53]: in both later cases the strongest effect was the increase of Voc: from 0.1

V to 0.5 V in Poly-PV, and similar in small molecule PV. As shown at Fig.6 of comparison below: It reflects Voc related to increase of Fermi level in SWCNT with EDLC charging of CNT in all OPVs. On the contrary in Pero-PV the Voc do not change at all, but only shape of IV curve is changing at Voc, reflecting the strong change of series resistance, Rs (which is the slope of dV/dI at Voc point). This is a clear indication that the w.f. of electrode is not important for PS-PV, since the internal p-i-n junction quickly forms inside PS, as sketched in Fig X. this raises Fermi levels in n-doped part of PS layer adjusting it to Fermi level of electrode CNT, independently where this level is initially. So as in all PS PV devices, the Voc does not depend much on w.f. of electrode, once the internal p-i-n is formed by dynamic photopoling[26,54] or poling[55]. Therefore, the most important effect of Vg is the increase of photocurrent Isc due to better charge collection by n-doped ETL and lower series resistance Rs of the device becoming smaller with Vg increase causing increase of FF.

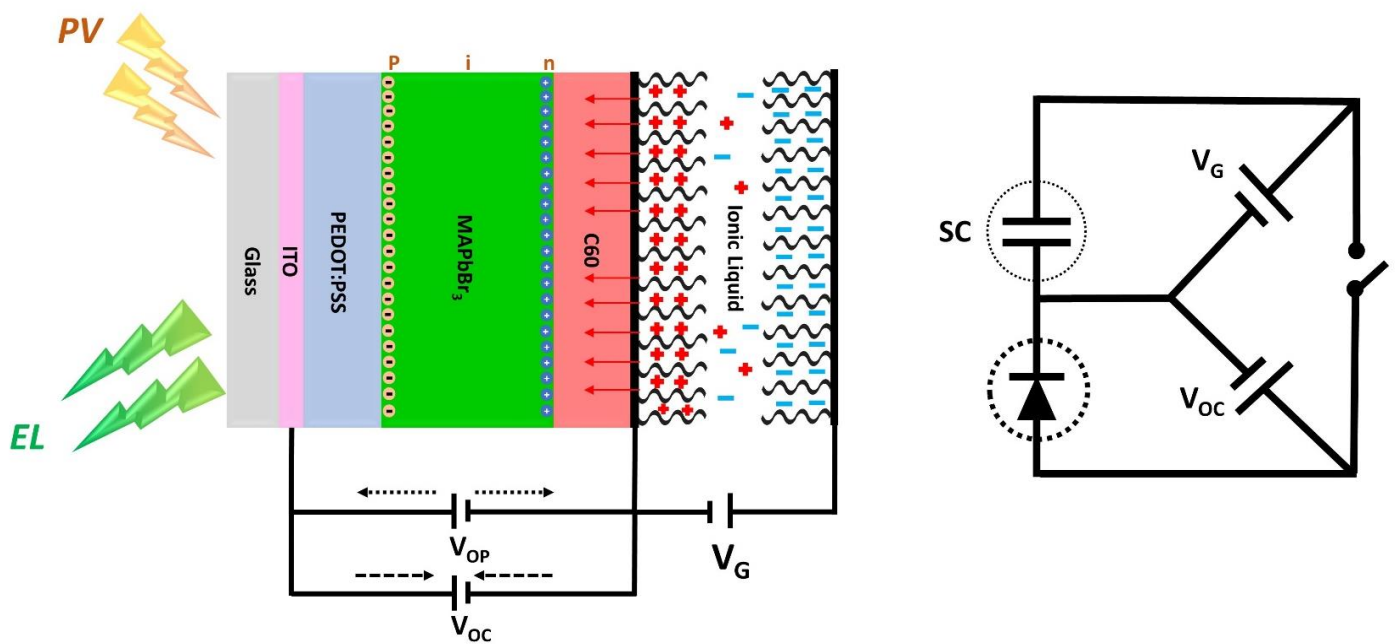

Figure 6. Schematic operation of PS-PV tuned by gating: a) at small Vg and upon light the PV operation is effective mainly by generating larger Isc with better FF, upon Vg raise, while Voc is not changing at all, since Vg is determined by inner p-i-n junction in PS, shown as intrinsic ions accumulated in PS at interfaces. Larger Vg> 2 V causes strong doping of ETL into n+ by plus ions of Ionic liquid DEME+ accumulation in C60 and lowering both barriers at C60/CNT and PS/C60 interfaces. However this processes do not effect much neither Isc, nor FF, since charge collection cannot be further much improved.  B) the equivalent circuit of gated PS-PV showing a supercapacitor SC on top of ionically tunable photodiode I-PV. Charging of SC can be enhanced by self-charging by Voc, or even solely done by Voc, which is increased by PIN formation in PS layer.

For comparison, we present the results of IV curves change with Vg for a similar Vg ranges in OPV (as originally presented in work of Voroshilov and co – authors[53]): S-shape changes to good FF even with Vg=0, due to polarization by EDL formation at interface (as in our case). Then further doping at higher Vg increases Voc due to Fermi level raise of SWCNT electrode, till the CB of ETL, when the conductivity of SWCNT.

So, the great and unusual observation is that $V_{oc}$ is not increased at all, although the Fermi level is known to raise in SWCNTs. This means that w.f. of SWCNT is not important for $V_{oc}$, it already was high due to PS physics. And this physics is now showing its nature in this gating difference.

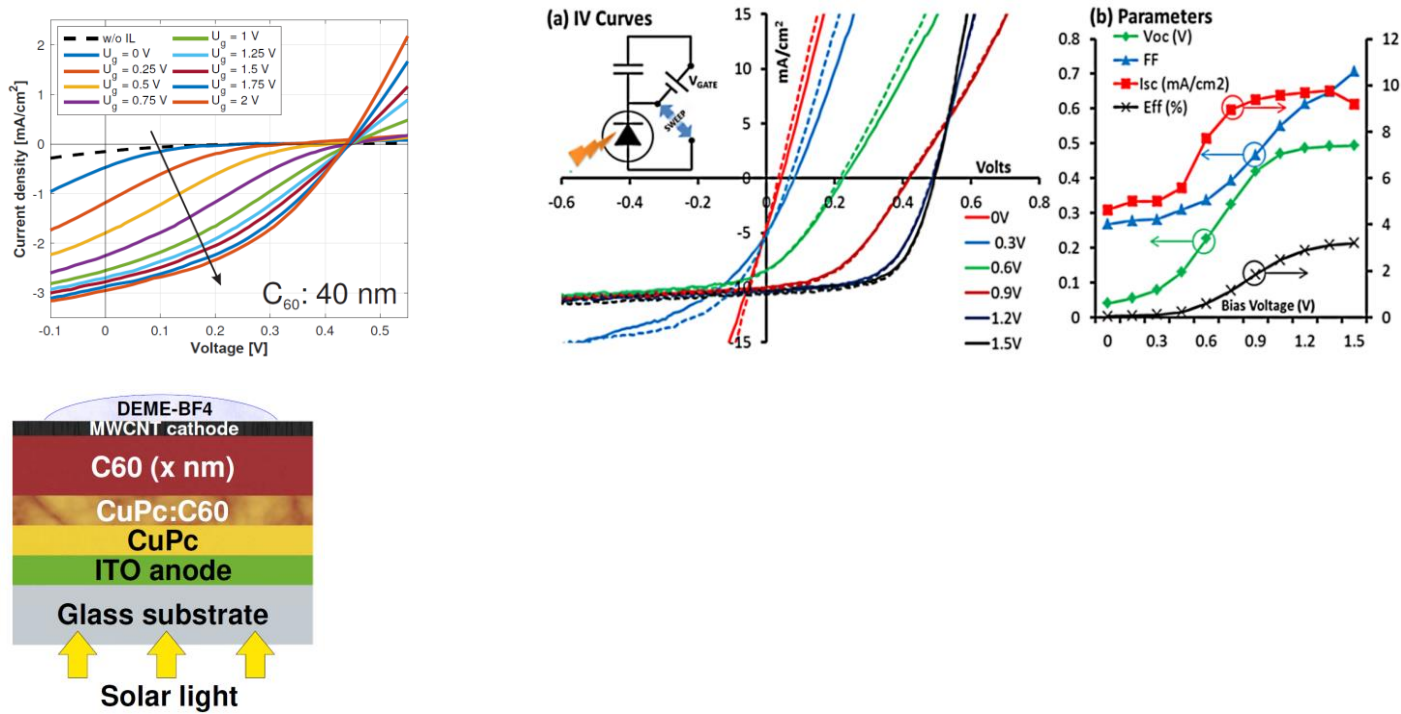

Fig. 7. IV curves of two typical OPV upon Ionic gating, demonstrating strong increase of Voc and Isc

In OPV the main effect is the dramatic change of Voc as shown in Fig. Y for two types of OPV: PHT/PCBM bulk heterojunction (from the work of Cook[56] and CuPc/C60 small molecule OPV (from work with ionic gating for small molecule OPV)[53].

This insensitivity of Voc on ionic gate Vg in PS-PV is an indirect prove that in PS the inner pin junction is formed, via the accumulation of its intrinsic ionic species, as shown at Fig. 7 above. In case of PS-PV the strong increase of Isc is due to improved collection of carriers by the N-doped ETL, and partially by lowering the series Rs.

The amazing second difference of PS-PV from OPV is the insensitivity of operation on Vg at higher Vg > 2.0 – 25 V. this is again can be understood as due to n+ heavy doping of C60.C70 at higher Vg, which only slightly increase charge collection, and do not degrade the PV operation, like in OPV, as observed by Saranin[56] in gated OPV tandems at high Vg. The interplay of inner p-i-n junction in PS layer with ions coming from ETL at higher Vg is an interesting challenge, that needs further investigation.

Mainly C70 and C60 ETL based devices have different levels of current generation and output performance response to gate voltage in dependence to thickness. At initial conditions, devices with 200 nm C70 based ETL have 1.5 times bigger short current density ~16 mA/cm$^2$ in comparison to 12 mA/cm$^2$ of 200 nm C60 based ETL cells (as it shown on Figure 2(a)(c)). The difference of J$_{sc}$ values is determined to higher absorption of C70 films in 450–600 nm visible range of light, and gives its contribution in photon harvesting and larger photocurrent due the larger quantity of generated excitons, as shown in works with OPV devices and planar PSC.[57,58] With an increase in the thickness of the electron transport layer from 200 to 300 nm, a difference between J$_{sc}$ performance of C60 and C70 devices becomes more obvious. Different gate bias response of devices is explained due energy structure, surface state, and interfaces between fullerene ETL, SWCNT cathode, and induced positive ions-DEME$^+$. First, a significant drop of series resistance was observed after deposition of ionic liquid on the cathode surface without applied gate bias. In turn, decreasing the contact resistance corresponds to doping effects, which occur at room temperature between the SWCNT-fullerene-IL interfaces. Carbon nanotubes-IL interfaces can have strong interaction via Van-Der-Waals bonding, $\pi$-$\pi$ bonding, and sidewall adsorption, as it presented in several reviews with modelling and experimental results[59,60].

At the same time, fullerene ETL-IL interface has another interaction in physics. Theoretical calculations, provided by Chaban and co-workers,[61] show that a very significant polarization effect appears between the room temperature ionic liquid and the C60 molecule. With use of the hybrid density functional theory (HDFT) and powered Born-Oppenheimer molecular dynamics (BOMD) simulations, authors showed that ion adsorption at C60 surface acquires systematically positive electrostatic charges 0.1–0.2e with imidazolium IL. Moreover, conduction and valence band orbitals are shared between fullerene and ions of ionic liquid as result of polarizing action. Semiconductor energy levels can be shifted up or down in the presence of different anions of ionic liquid (authors showed LUMO raise with Cl$^-$, NO$_3^-$ ions and drop with PF$_6^-$ anion), with band gap tuning. Therefore, we suggest, that initial contact between C60/C70 ETL with BF$_4^-$ anion shifted down LUMO with improving of energy level matching with SWCNT W$_f$. As result, such polarizing action can give contribution to initial drop of contact resistance between ETL and cathode during IL cathode soaking. Applying this model to our case with three-side interface, direct charge transfer can be done in accumulation regime at V$_{gate}$≤2.00 V, when ETL LUMO and SWCNT W$_f$ have equal energy level. As It shown in JV gate dependence of ETL film diode structure, asymmetry of forward reverse curve corresponds to n- type accumulation in depth of ETL with sequential decreasing of differential resistance and rectification of JV curve at V$_{gate}$>2.00 V. Thereby, induced i-n junction and expansion of n-doped area in ETL with increasing V$_{gate}$ have improved charge collection and current generation due lower LUMO position and higher concentration of carriers. Advantage of junction in transport layers for PSC was presented already in a work by Jung at al.,[62] where authors developed n-i-p Spiro-Ometad HTL with co-evaporation doping methods. N-type doped spiro-OMeTAD was adjusted to perovskite HOMO for efficient hole extraction while a p-type layer formed optimal level matching with the gold electrode. Response of output parameters from gate bias for C60 and C70 ETL devices caused by different energy positions of LUMO levels and molecule arrangement with different carbon bonds curvature. It is well known that C70 have lower a position in comparison to C60, approximately 4.2 eV to 3.9 eV[57,63], hence, C70 ETL initially has better alignment with SWCNT $W_f$ and lower potential losses. Consequently, C60 ETL requires a higher-level pf n- type accumulation for level matching with cathode $W_f$ and electron extraction. C60 molecule is buckminsterfullerene or buckyball, and its packing changes gradually from hexagonal to a cubic form, while C70 elongated fullerenes have a rugby ball-like shape, and prefer the hexagonal packing[64]. Different surface curvature due to chemical bonding and molecule packing allows for specific ion adsorption and distribution under gate bias, which influences the depth of charges accumulation.

**Conclusions**

In summary, we demonstrated a combined approach of interfacial engineering and n-type accumulation in advanced planar p-i-n perovskite solar cell with ionic gate and ultra-thick ETL based on fullerene. Operation capability of such thick 200–300 nm fullerene-based ETL is presented for the first time for perovskite solar cells. Initial high series resistance of SWCNT with thick ≥200 nm ETL interface was dropped from extremely high kOhms*cm$^2$ values to ≤20 Ohm*cm$^2$ due to dipolar polarization by $DEME^+$ and $BF_4^-$ ions and Vg gate bias induced n-type doping of CNT/ETL via injected carriers. Sequential increase of gated bias successfully transformed SWCNT-fullerene Shottky contact to ohmic junction while simultaneous initial intrinsic and high ohmic nature thick 200–300 nm C60/C70 fullerene films were compensated by charge accumulation in ETL depths. Appearance of i-n junction (induced by gate bias) was confirmed by diode structure, JV measurements with presence of rectification ratio at $V_{gate}$ ≥ 1.5 V.

The different response of JV performance to gate bias and ETL thickness was observed for C60 and C70 devices. This effect occurs due to the higher curvature of C70 molecules packing that allows better IL cations-anions adsorption and distribution under the applied field, and initially lower LUMO position, which provides matching SWCNT $W_f$ at lower gate bias. Therefore, C70 ETL have lower spread of $J_{sc}$ in dependence to thickness with increased $V_{gate}$, while C60 ETL devices have brighter expression of dependence to layer thickness.

Finally, devices showed dramatic improvement of output parameters with changes starting from the S-shape JV curve to 11+ % PCE performance. Best efficiencies were demonstrated on 200 nm C60 cells and 250 nm C70 devices at $V_{gate}$ = 2.25 V and 2.50 V, respectively. For C60, most performing cells $J_{sc}$ gain was + 54.2%, FF increased in 2.7 times, and efficiency grew from 2.46% to 11.29%, respectively, and for C70, $J_{sc}$ improved +44 %, FF increased in ~3 times, and efficiency grew from 1.47% to 11.13%.

Using ionic gate for cathode junction opens new perspectives for improving electron transport materials of various thicknesses and promising materials for electrode application with ambient, vacuum-free processing.

**Acknowledgements**

Support for this work was provided from the Ministry of Education and Science of the Russian Federation in the framework of Increase Competitiveness Program of NUST "MISiS" (No. K2-2019–014). Partial financial support by Welch grant AT-1617 is highly appreciated.